\documentclass{nrlreport} 
\usepackage[section]{placeins}
\usepackage{amssymb}
\usepackage{gensymb}

\defaultbibliography{PVDF_Memorandum_Report.bib}

\nrlheadertitle{3D Printed PVDF}
\nrlheaderauthor{Alec Ikei [et al.]}
\nrlbranch{Acoustic Signal Processing and Systems Branch\\Acoustics Division}

\nrlauthortwo{Gregory Yesner (Current Postdoc)}   \nrlbranchtwo{Physical Acoustics Branch\\Acoustics Division}

\nrlreportnumber{MR}{7165}{20-10,123}
\nrlapprovaldate{September 30, 2020}

\nrlwarning{}  \nrldistribution{Distribution A: Approved for public release: distribution unlimited}
\nrlclassifiedby{}  \nrldeclassifyon{}  \nrldissemination{}

\nrlreporttype{NRL Memorandum Report}
\nrldatescovered{02/2018 - 02/2020}  \nrlauthorlist{Alec Ikei, James Wissman, Charles Rohde, and Gregory Yesner}

\nrlcontractnumber{}  \nrlgrantnumber{}  \nrlprogramelement{NISE}
\nrlprojectnumber{}  \nrltasknumber{}  \nrlworkunitnumber{N2N8}

\nrlorganizationaddress{Naval Research Laboratory\\4555 Overlook Avenue, SW\\Washington, DC 20375-5320}  
\nrlsponsoraddress{Naval Research Laboratory\\4555 Overlook Avenue, SW\\Washington, DC 20375-5320}
\nrlsponsoracronym{NRL-NISE}  \nrlsponsorreportnumber{}
\nrlsupplementarynotes{Karles Fellowship}  \nrlsubjectterms{3D-printed piezoelectric polymers, PVDF fabrication, Corona poling}

\nrlreportclass{Unclassified Unlimited}  \nrlabstractclass{Unclassified Unlimited}  \nrlthispageclass{Unclassified Unlimited}
\nrlabstractlimitation{Unclassified Unlimited}  \nrlresponsibleperson{Alec Ikei}  \nrltelephonenumber{202-404-4816}

\begin{document}
\title{3D Printed PVDF}

\author{Alec Ikei\\James Wissman (Former Postdoc)\\Charles Rohde}

\date{\today} 
\maketitle    

\nrlabstract{In this paper we report on the 3D printing and testing of the piezoelectric polymer polyvinylidene difluoride (PVDF).  Samples of PVDF were fabricated using a fused deposition modeling (FDM) 3D printer and then activated using a corona poling process.  The d33 piezoelectric coefficient, which is related to the overall piezoelectric performance, was experimentally measured using a d33 meter to be 6 pC/N.  While less than commercially available PVDF fabricated using traditional techniques (which can have a d33 of 10 – 40 pC/N), the value of 6 pC/N achieved in this work is several orders of magnitude larger than comparable previously published results for 3D-printed PVDF, and as a result represents a significant step in the 3D printing of piezoelectric polymers.} 

\tableofcontents  
\listoffigures    
\listoftables     

\begin{executivesummary}
The purpose of this project is to leverage 3D printing to produce complex structures for the fabrication of acoustic metamaterials.  To accomplish this goal, this project’s approach is to 3D print piezoelectric devices.
PVDF (polyvinylidene difluoride) was chosen for its relatively strong piezoelectric response (10-40 \units{\frac{pC}{N}}) when prepared properly, as well as its relatively high resistance to corrosion and environmental friendliness when compared to ceramics like PZT (lead zirconate titanate).  Another key feature is that it also possible to print it on an FDM (fused deposition modeling) based 3D printer.  To prepare it properly, a sufficient amount of shear stress must be applied while heating and applying a large electrical field across the material.  The shear stress and electric field serve to align the polymer chains and electric dipoles in a particular direction, while the heat provides mobility to do so.  The approach used in this work applies high electrical potential difference between the print head and the bed of the FDM machine in-situ to accommodate these needs.  In this work, 3D printed PVDF that was stretched and poled post-process was shown to exhibit 6\units{\frac{pC}{N}}, performing better than other efforts of FDM manufacture of PVDF.
\end{executivesummary}

\chapter{Background}
\section{Piezoelectric Devices}
Piezoelectric devices are commonly used in acoustic applications as transducers and sensors.  The piezoelectric phenomenon converts strain into a potential difference between the surfaces of the sample.  This relationship is defined in Eq. \ref{eqn.tensor}, where \textbf{X} is strain, \textbf{d} is the piezoelectric coefficient, \textbf{E} is the electric field, and \textbf{M} is the electrostrictive coefficient \cite{Uchino-2009}.  The d tensor is compressed into a second rank tensor through symmetry arguments \cite{Uchino-2009} and the non-linear electrostrictive contribution, present in all materials, is small in piezoelectric materials at low fields (Eq. \ref{eqn.2ndrank}) \cite{Newnham-2005}.  Additionally, convention dictates that the ``1'' direction refers to the stretch direction and the ``3'' direction refers to the polarization direction.  As an example, this means that if a sample is pulled in the 1 direction, the electric response in the 3 direction is related by the d$_{31}$ component.  

\begin{equation}
X_{kl}=\sum_i d_{ikl}E_i+\sum_{i,j}M_{ijkl}E_iE_j
\label{eqn.tensor}
\end{equation}

\begin{equation}
X_j=\sum_id_{ij}E_i
\label{eqn.2ndrank}
\end{equation}

\section{PVDF Manufacturing}
In many applications, ceramic piezoelectrics have been favored due to their better piezoelectric performance when compared to PVDF (polyvinylidene difluoride) (PZT 501A and PVDF have a d$_{33}$ value of 400 \units{\frac{pC}{N}} and 20\units{\frac{pC}{N}} respectively \cite{Uchino-2009}).  However, PVDF is printable through FDM (fused deposition modeling), does not contain lead, is corrosion resistant, relatively cheap and is not brittle.  

\begin{figure}
\includegraphics[width=0.4\textwidth]{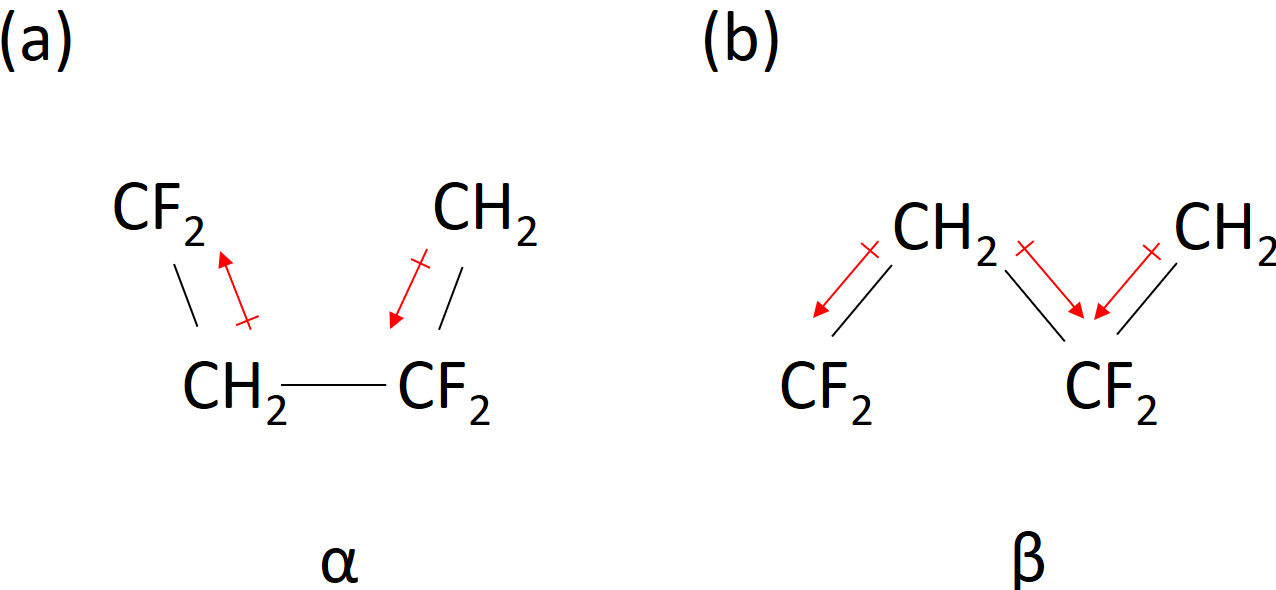}
\caption{(a) Monomer of $\alpha$ phase of PVDF.  Net dipole moment transverse to carbon chain direction is zero.  (b) Monomer of $\beta$ phase of PVDF.  Non-zero dipole moment transverse to the carbon chain, which is necessary for poling.}
\label{fig.alphavsbeta}
\end{figure}

PVDF is a fluorinated hydrocarbon, and has either two hydrogen or two fluorine atoms connected to each interior carbon atom (CF$_{2}$CH$_{2}$)$_{n}$, where the n denotes that the elements inside the parenthesis are repeated to comprise the polymer.  It has several phases that refer to the orientation of its molecular structure, which are the basis of its piezoelectric properties.  If the fluorine atoms and hydrogen atoms are on separate sides of the chain, it creates a dipole moment for each monomer and is called the $\beta$ phase \cite{Uchino-2009}.  When cooled from the liquid phase, it forms the $\alpha$ phase, where the dipoles within the monomer oppose each other and cancel out, which makes each monomer and the polymer non-polar (Figure \ref{fig.alphavsbeta}).  Therefore, it must be mechanically stretched to convert it to the polar $\beta$ phase to prepare for poling \cite{Uchino-2009} \cite{Gomes-2010}.  In the poling step, an electric field is applied to the $\beta$ phase PVDF to orient the dipoles in bulk.  When strain is applied to the crystal it deforms the molecular structure, bringing together or pushing apart the dipoles of the monomers.  This changes the density of dipoles and is observed macroscopically as a development of electric potential between the crystal's surfaces i.e. the piezoelectric effect.  

Many commercially produced piezoelectric PVDF films are made by extruding it in a relatively thick sheet, which is then stretched to around 400-500\%  of its original length\cite{Lee-2014}, electroded and corona poled.  The commercial sample used in this work is a product of this process (Component Distributors Inc., part \#3-1003702-7).  While mechanical stretching is the most common, other methods of inducing $\beta$ phase content exist, such as high pressure treatment, annealing, ultra-rapid cooling, electrospinning, or crystallization from solution \cite{Lederle-2020}.  

\section{3D Printed Sensors}
3D printing has become popular as a way to rapidly prototype and create complicated designs, because it does not rely on machining material away from parts.  The rapid prototyping makes the design process quicker, smoother and less expensive.  Incorporating active elements into 3D printed designs allows the sensor design process to reap these benefits.  One application of interest is in 3D printed acoustic sensors.  Acoustic sensors have been 3D printed that sense through bionic, capacitive, or piezo-resistance methods \cite{Xu-2017}.  

3D printers have also been used to print piezoelectric devices.  Kim et. al implemented a way to 3D print piezoelectric films using corona poling to induce piezoelectric structure \cite{Kim-2017}.  This process relies on the shear from printing to transform the melt into the $\beta$ phase and applies high voltage to the print head and scans over the layer after it has been printed.  The voltage is high enough to ionize the air surrounding the nozzle (corona field).  The electric field of the ionized particles deposited on the surface of the print and the grounded base plate beneath the printed layer induce the dipole orientation necessary for piezoelectric response in PVDF.  The print was determined to have a d$_{33}$ value of 4.8 x 10$^{-2}$ \units{\frac{pC}{N}}.  While Kim et. al rely on a corona field to pole their print, in this work the voltage was applied directly to the print head to supply the necessary electric field.  Additionally, with post process poling the sample was able to achieve a higher piezoelectric response of 6\units{\frac{pC}{N}}.

The mechanical compliance of PVDF has made it a good material for 3D printed soft robotics.  3D printed soft robots have also used other polymers, hydrogels and elastomers and have been actuated through thermal, electrical, pH-based, light-based and magnetic means \cite{Ali-2016}.  Soft robotics are good at handling delicate materials and conforming to space restrictions, which make them useful in many industrial, medical and human interaction settings.

\chapter{Results}
\section{Additive Manufacturing}
\subsection{Creating PVDF FDM Filament} 
Kynar 740 PVDF pellets and a Filabot EX2 filament extruder were purchased to create the filament.  Through control of the extrusion parameters (temperature, extrusion speed, cooling, drawing speed), the amount of air bubbles was decreased, and cross sectional geometry and consistency of the filament diameter was adjusted to create filament that was suitable for 3D printing in an Ultimaker 3.  Even while keeping all the parameters consistent, however, the inconsistencies of the extruder’s motor resulted in it over/under extruding.  A paper shredder was used to reprocess over/undersized filament that was not within the diameter specifications of the printer. 
Although the melting point of PVDF is approximately 177$\degree$C, extruding at this temperature it appears to cause additional air bubbles and diameter inconsistency.  In general, best results were achieved when extruding in the 200-220$\degree$C range.

\begin{figure}
\includegraphics[scale=.8]{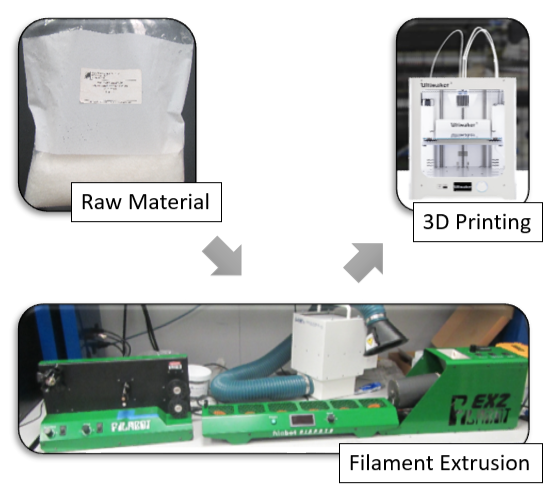}
\caption{PVDF pellets (Raw Material) are processed into filament (Filament Extrusion), which is then used as feed material in 3D printers.}
\end{figure}

\FloatBarrier
\subsection{3D Printing with PVDF}
\begin{figure}
\includegraphics[scale=.4]{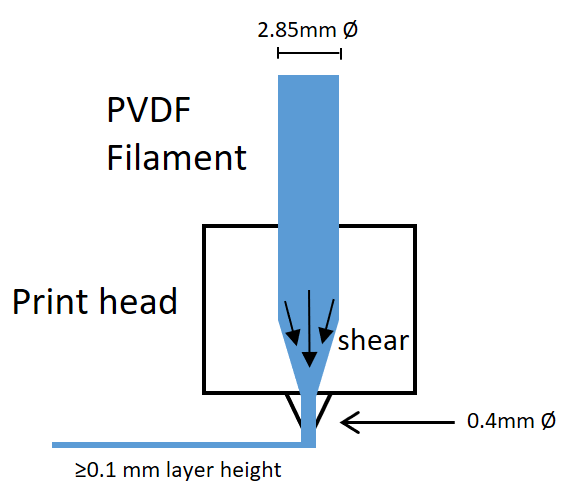}
\caption{Shear occurs through restriction of the feed material's cross-sectional area in the print head, and during deposition into thin layers.}
\label{fig.shear}
\end{figure}

\begin{figure}
\includegraphics[scale=.4]{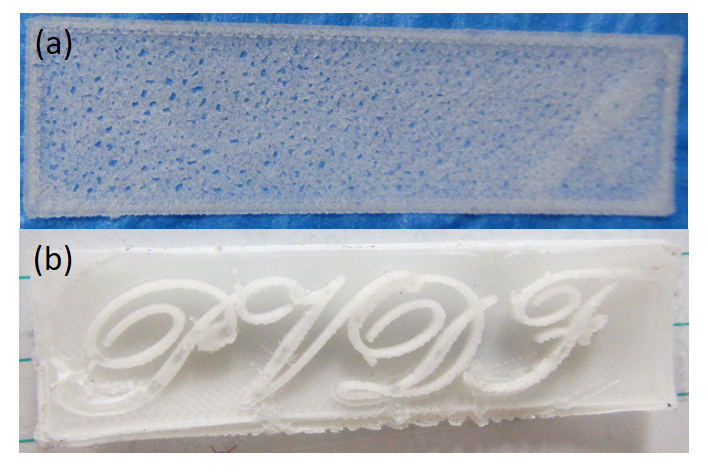}
\caption{Printed PVDF. (a) While the overall geometry was correct, there was significant porosity in the print. (b) Printed PVDF part, with "PVDF" embossed on the surface.}
\label{fig.porous}
\end{figure}

In the printing setup, Cura (v2.2.6) was used to slice the models into gcode (movement code), which was run on an Ultimaker 3.  FDM type 3D printers, like the Ultimaker 3, drive filament into a print head, which heats and extrudes it to build a model layer by layer.  To apply the shear stress and required heat, thin layers of PVDF were printed using a small diameter nozzle (0.4mm) (Figure \ref{fig.shear}).  Tuning the printer parameters was a crucial step.  In initial attempts to print, the PVDF extruded from the nozzle would roll up into a ball or not extrude from the nozzle at all.  After troubleshooting from the failed prints (e.g. Figure \ref{fig.porous} (a)), the sample in Figure \ref{fig.porous} (b) was printed by modifying following settings from the pre-set “Extra Fine” printing profile, listed in Table \ref{tab.parameters}.  The settings log is listed in the appendix \ref{app.one}.

\begin{table}
\centering
\caption{Print parameter deviations from "Extra Fine" printing profile in Cura}
\label{tab.parameters}
\begin{tabular}{|c|c|}
\hline
Print Temperature & 220\degree C-240\degree C\\ \hline
Infill Percentage & 100\\ \hline
Line Width & 0.3mm\\ \hline
Material Diameter & 2.55mm\\ \hline
Layer Thickness & 0.1mm-0.15mm\\ \hline
\end{tabular}
\end{table}

\FloatBarrier
\subsection{Multi-Material Printing}
\begin{figure}
\includegraphics[scale=1]{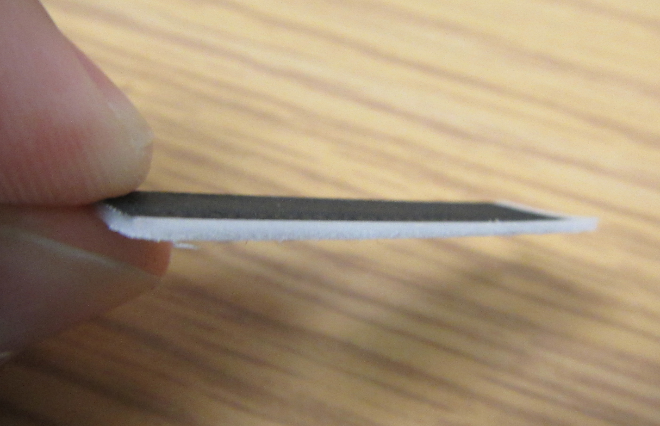}
\caption{PVDF layer on the bottom with carbon PLA layer on top}
\label{fig.dual layers}
\end{figure}

\begin{figure}
\includegraphics[scale=.5]{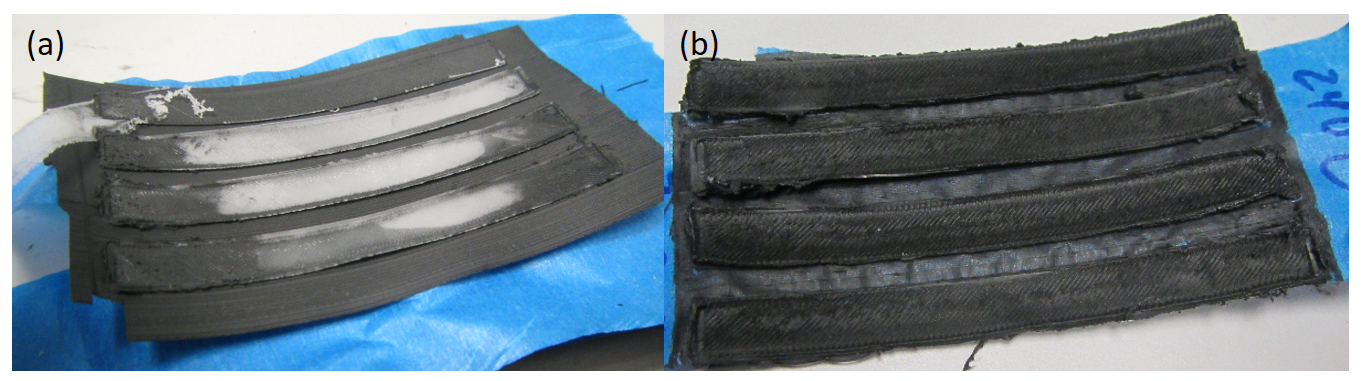}
\caption{(a) An attempt to print a carbon PLA raft, with PVDF and a carbon PLA layer on top.  There was significant delamination of the PVDF layer and the carbon PLA layer. (b) Another attempt with different parameters.  Intact lamination of all layers, but with visible warping.}
\label{fig.plaraft}
\end{figure}

To create an electrode for evenly applying a potential across the surface of the PVDF, layers of carbon infused PLA (polylactic acid) were printed above and below the PVDF layer.  To start off, PVDF was printed directly onto the build plate, with carbon PLA on top in Figure \ref{fig.dual layers}.  However, it was significantly more difficult to make a print with a layer of carbon PLA on the bottom.  This was due to the PVDF peeling off the carbon PLA during the print, which was a result of thermal contraction and poor bonding between the two material types (Figure \ref{fig.plaraft}).  Including additional PLA around the bottom layer (i.e. adding a raft), as well as increasing print temperature and lowering the printing speed improved its ability to stay relatively flat and successfully print the three layers, using the parameters from Table 1.

\FloatBarrier
\section{Piezoelectric Poling and Benchmarking}
\subsection{Oil Bath Poling}
\begin{figure}
\includegraphics[scale=.5]{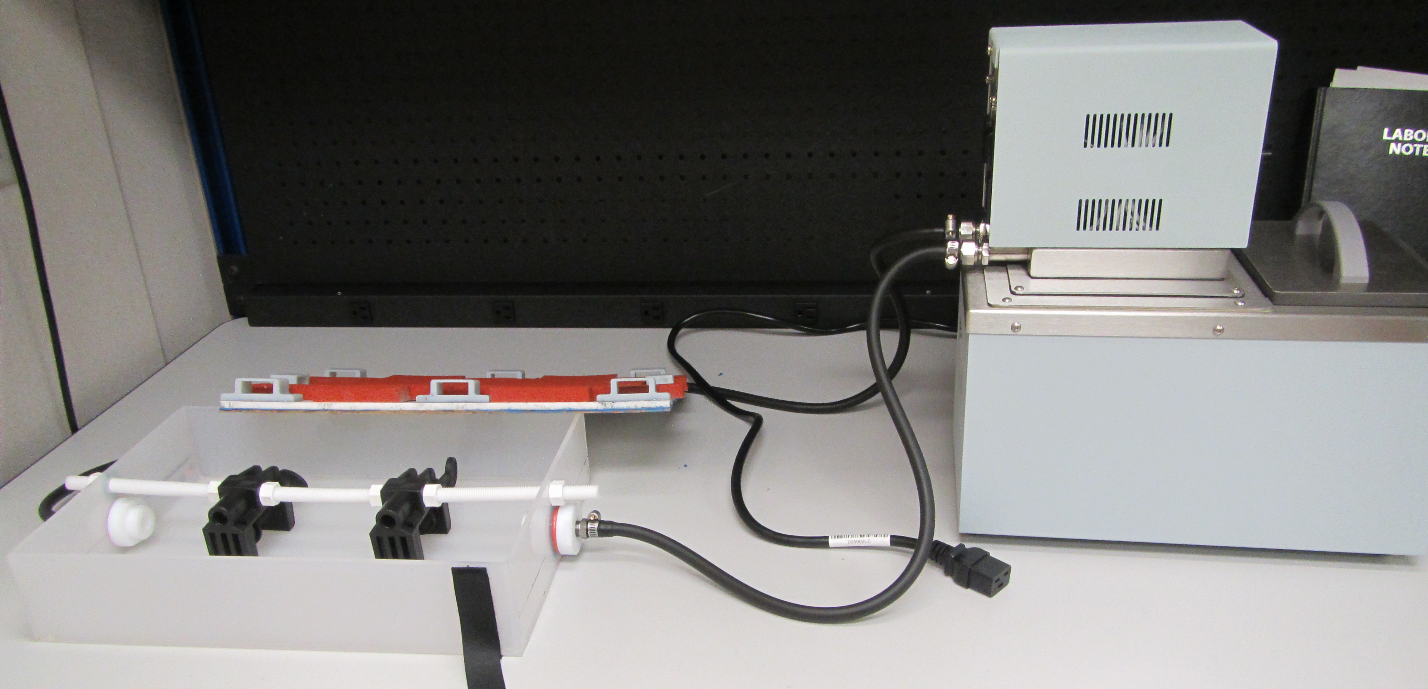}
\caption{Oil bath with lid and heated fluid circulator.  Bath constructed out of polypropylene, with clamps and fittings made out of ABS and PTFE (Teflon) respectively.}
\label{fig.oil}
\end{figure}

In demonstration of the feasibility of poling 3D printed PVDF with contact electrodes, a post-process poling setup was devised.  A heated, circulating fluid pump was purchased to provide silicone oil at a constant temperature in a dielectric bath.  The dielectric bath was constructed out of polypropylene, with the clamps and fittings chosen to provide additional electrical insulation (Figure \ref{fig.oil}).
With this heat bath, poling of a PVDF sample was attempted, but it did not yield a measurable piezoelectric coefficient.  This was due to early dielectric breakdown of the PVDF layers, which was likely caused by local variation in the print thickness and squeezing from the thermal expansion of the clamps.

\subsection{Piezoelectric Coefficient Testing}
In preparation of future needs to characterize piezoelectric films, a cantilever test was performed, using commercially supplied piezoelectric sheets (Component Distributors Inc., part \#3-1003702-7) that were bonded using epoxy (Devcon, part \#14250).  A cantilever test measures the tip deflection of a cantilevered sample due to the electric field applied between its top and bottom surfaces.  The strain is calculated from the tip deflection of the sample and can be used to determine its d$_{31}$ value.  The tip deflection was measured using a Laser Doppler Vibrometer (LDV) (CLV-2534, Polytec) with its data acquired by an oscilloscope.  The piezoelectric sheets consisted of a 110$\mu$m thick PVDF layer sputtered with 700 Å of copper and 100 Å of nickel on both sides.  The sample measured 82.87mm in length.

\begin{figure}
\includegraphics[scale=.55]{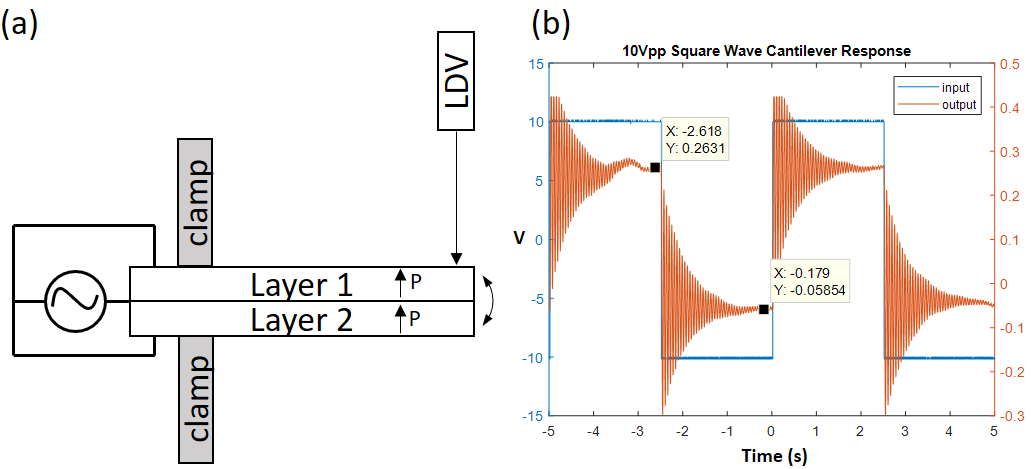}
\caption{(a) Parallel bimorph setup for LDV cantilever deflection measurement, where polarization is in the same direction.  (b) Input voltage (blue) vs cantilever actuation (orange).}
\label{fig.cantilever}
\end{figure}

From Figure \ref{fig.cantilever}b, the difference in the LDV output voltage indicates that the actuation distance was 32$\mu$m, as calculated through Eq. \ref{eqn.parallel cantilever}.  In the setup shown in Figure \ref{fig.cantilever}, the appropriate equation for a parallel configuration bimorph is described in Eq. \ref{eqn.parallel cantilever}, where $\delta$ is tip deflection, $d_{31}$ is the piezoelectric strain coefficient, L is length, t is thickness of the bimorph (i.e. twice the sheet/layer thickness) and V is applied voltage \cite{Uchino-2009}.  The bimorph consisted of two strips of the piezoelectric film epoxied together with their polarization oriented in same direction (parallel configuration).  This gives twice the actuation for the sample applied voltage in comparison to the antiparallel configuration, where the strip polarization are oriented in the opposite direction\cite{Uchino-2009}.  0.32V corresponds to a $d_{31}$ value of 15\units{\frac{pC}{N}}.  The manufacturer specifies a piezoelectric coefficient of 23 \units{\frac{pm}{V}}, which is equivalent to 23\units{\frac{pC}{N}}.

\begin{equation}
\delta =3d_{31}(L^2/t^2)V
\label{eqn.parallel cantilever}
\end{equation}

\FloatBarrier
\subsection{Corona Poling}
Local variation in the thickness of the print lead to early dielectric breakdown through the PVDF in the oil bath, so to verify that it was possible to pole the samples that were 3D printed on the Ultimaker 3, a corona poling method was designed and made to combat the issues previously encountered.  In this setup, a grounded metal plate was placed under an acrylic enclosure to prevent external leakage of charge.  A high voltage power supply (PS300, SRS) was used to apply up to 20kV to a tungsten needle, placed at the top of the enclosure.  A grounded piece of metal was placed under the needle to direct the ionized air towards the sample.

\begin{figure}
\includegraphics[height=.5\columnwidth]{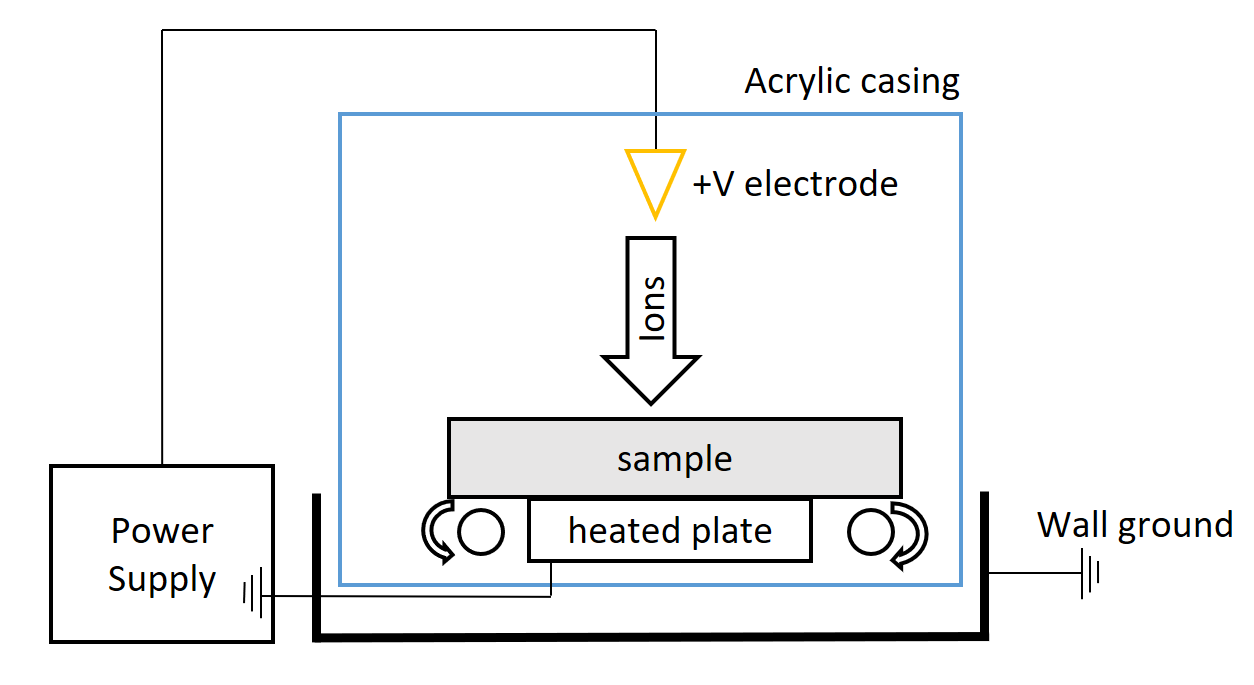}
\caption{Corona poling setup.  High voltage was applied to a tungsten needle to ionize the air.  The PVDF was stretched while under high temperature and high electrical fields.}
\label{fig.corona}
\end{figure}

Many corona poling setups also apply elevated temperatures and high levels of strain to the PVDF film, to lower the coercive field strength and align the polymer backbone respectively (the coercive field is the electric field needed to reverse the piezoelectric polarization of the material).  To do this to the printed parts, a heater and film stretcher were constructed mostly out of parts from an old 3D printer (Figure \ref{fig.corona}).  The heating element was placed in an aluminum block and a thermistor provided the feedback control to keep the block at a set temperature.  The stretching mechanism utilized a 3D printed worm gear and a small stepper motor, operated by a control board from a 3D printer to apply 400-500\% strain at 80$\degree$C.  Using this setup, a 3D printed piezoelectric film sample with a d$_{33}$ value of 6\units{\frac{pC}{N}} was achieved.  The d$_{33}$ measurement was taken by a d$_{33}$ meter a day after the poling process took place, to ensure that the remnant polarization was measured.

\FloatBarrier
\section{Simultaneous Poling and Printing}
\subsection{High Voltage FDM}
As mentioned in the background, it is necessary to apply strain, high electric fields and elevated temperatures to convert $\alpha$ phase PVDF to its piezoelectric form.  It has been shown that it is possible to do these processes in series in Kim et al., but the current work was done to do these processes simultaneously.  To reduce the costs involved if the printer were to break from having high voltage applied, an inexpensive FDM 3D printer (Ender 3) was purchased.  A steel plate was placed on the print bed to serve as a high voltage electrode.  The ground lead was attached to the top of the print head.  

\begin{figure}
\includegraphics[scale=.7]{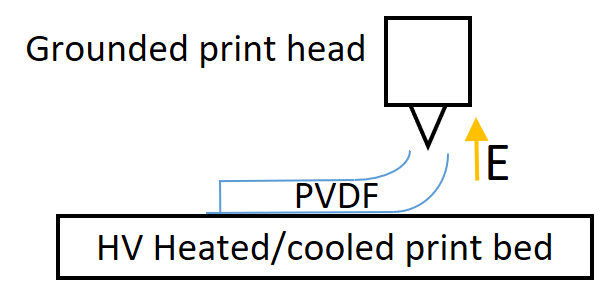}
\caption{In-situ poling setup.  Print bed is charged instead of print head to reduce risk of damage to the printer.}
\end{figure}

In normal use, the print bed of a 3D printer is heated to around the glass transition temperature (around 55\degree C for PLA).  However, the print bed was instead cooled using liquid nitrogen, dried and placed back on the printer.  This was done to help set in place the dipole alignment and material phase through the rapid cooling of the sample.  Making the print bed extremely cold mitigates the chance of reversion to the unpoled state due to retained heat.  This method has not yet yielded a measurable piezoelectric coefficient in any samples, but further work is planned to improve these results, which is noted in the summary section below.  

\FloatBarrier
\chapter{SUMMARY}
The ability to take pellets of PVDF stock, create a workable filament, and print it with conductive material has been shown.  While early attempts to extrude PVDF proved more challenging than extruding with more conventional materials like PLA, tuning the extrusion parameters based on failed filament and failed printed parts improved further attempts.  Multi material samples were printed, which would enable fully printing sources and sensors.

To characterize piezoelectrics, the performance of a bimorph cantilever can be used to determine the $d_{31}$ piezoelectric coefficient.  The setup in this work used an LDV to measure the deflection and a waveform generator to supply voltage to the bimorph.  This was performed on commercially obtained PVDF sheets, which showed that the commercial sheet had 68\% of the expected piezoelectric value.

Both a heated dielectric bath and a corona poling setup were made to pole printed parts ex-situ.  While the heated bath proved to be unreliable, the corona poling showed that the printed and stretched parts were able to reach d$_{33}$ values of 6\units{\frac{pC}{N}}.  A high voltage in-situ printing setup was made, but has not yet yielded a measurable piezoelectric coefficient.  In future work, a PVDF copolymer (PVDF-TrFE) will be used instead as the starting stock.  This co-polymer forms $\beta$ phase during simultaneous heating and poling\cite{Pi-2014}, without the use of mechanical stretching.  This would eliminate the shear force requirement, and may produce a better result when printing and poling simultaneously. 

\bibliography{PVDF_Memorandum_Report}

\begin{bibunit}
\putbib
\end{bibunit}

\appendix

\chapter{Appendix One}
\label{app.one}
Example Cura Profile Settings for PVDF Extrusion
\begin{verbatim}
ultimaker3_pvdf[general]
version = 3
name = pvdf
definition = ultimaker3

[metadata]
setting_version = 4
quality_type = fast
type = quality_changes

[values]
adhesion_extruder_nr = 0
adhesion_type = skirt
default_material_bed_temperature = 100
layer_height_0 = 0.15

ultimaker3_extruder_right_#2_pvdf[general]
version = 3
name = pvdf
definition = ultimaker3

[metadata]
setting_version = 4
position = 1
quality_type = fast
type = quality_changes

[values]

ultimaker3_extruder_left_#2_pvdf[general]
version = 3
name = pvdf
definition = ultimaker3

[metadata]
setting_version = 4
position = 0
quality_type = fast
type = quality_changes

[values]
cool_fan_speed = 0
cool_fan_speed_max = 0
default_material_print_temperature = 220
infill_angles = [90]
infill_sparse_density = 100
initial_layer_line_width_factor = 100
material_diameter = 2.7
skin_angles = [90]
speed_layer_0 = 5
speed_print = 5
wall_line_width_x = 0.35
\end{verbatim}

\end{document}